\documentclass[sigplan,screen]{acmart}
\AtBeginDocument{%
  \providecommand\BibTeX{{%
    \normalfont B\kern-0.5em{\scshape i\kern-0.25em b}\kern-0.8em\TeX}}}
\usepackage{setspace}

\usepackage{subcaption}
\usepackage{listings}
\usepackage{comment}
\usepackage{url}
\usepackage{amsmath,amsfonts}
\usepackage{bbm}
\usepackage[noend]{algpseudocode}
\usepackage{algorithm}
\usepackage{tcolorbox}
\usepackage{tikz}
\usepackage{float}
\usepackage[mode=buildnew]{standalone}
\usetikzlibrary{matrix,chains,positioning,decorations.pathreplacing,arrows}
\usepackage{pgfplots}
\pgfplotsset{compat=1.10}
\usetikzlibrary{shapes.geometric,arrows,fit,matrix,positioning, shapes, calc}
\usepackage{ulem}
\usepackage{multirow}
\pgfarrowsdeclarecombine{twotriang}{twotriang}
{stealth'}{stealth'}{stealth'}{stealth'}

\newcommand{\Ignore}[1]{}

\renewcommand{\baselinestretch}{0.96}

\begin{document}

\title{ESBMC-Solidity: An SMT-Based Model Checker for Solidity Smart Contracts}

\author{Kunjian Song}
\author{Nedas Matulevicius}
\affiliation{%
  \institution{University of Manchester, UK}
  \country{}
}
\email{kunjian.song@postgrad.manchester.ac.uk}
\email{nedas.matulevicius@postgrad.manchester.ac.uk}

\author{Eddie B. de Lima Filho}
\affiliation{%
  \institution{TPV Vision Innovator, Brazil}
  \country{}
}
\email{eddie_batista@yahoo.com.br}

\author{Lucas C. Cordeiro}
\affiliation{%
  \institution{University of Manchester, UK}
  \country{}
}
\email{lucas.cordeiro@manchester.ac.uk}
\renewcommand{\shortauthors}{Song et al.}

\begin{abstract}
Smart contracts written in Solidity are programs used in blockchain networks, such as Etherium, for performing transactions. However, as with any piece of software, they are prone to errors and may present vulnerabilities, which malicious attackers could then use. This paper proposes a solidity frontend for the efficient SMT-based context-bounded model checker (ESBMC), named ESBMC-Solidity, which provides a way of verifying such contracts with its framework. A benchmark suite with vulnerable smart contracts was also developed for evaluation and comparison with other verification tools. The experiments performed here showed that ESBMC-Solidity detected all vulnerabilities, was the fastest tool, and provided a counterexample for each benchmark. A demonstration is available at \url{https://youtu.be/3UH8_1QAVN0}. 
\end{abstract}




\maketitle

\section{Introduction}
\label{sec:intro}


The blockchain system is a distributed ledger technology that forms the primary mechanism behind Bitcoin, Ethereum, and alternative cryptocurrencies~\cite{bashir2020mastering}. It can be considered as a singly linked list of blocks~\cite{soloro2019hands}, where each of them contains a set of unmodifiable transactions. This way, such a technology serves as a distributed tamper-resistant record of such transactions~\cite{yaga2019blockchain}. Ethereum, for instance, can be regarded as a state machine whose global state is updated by those transactions, which indeed constitute state transitions. 

The transactions are performed by smart contracts, which are programs automatically executed on blockchain networks when specific conditions are met~\cite{bashir2020mastering}, which encode business logic. For instance, such conditions can be an exchange of cryptocurrency or even a process of content unlocking if a digital rights management system is involved. Indeed, transactions act as stimuli to smart contracts. Nevertheless, such contracts must first be written in a given language. In the case of Etherium, smart contracts are written in Solidity, which is an object-oriented language for programs to be run on the Ethereum virtual machine (EVM) ~\cite{soloro2019hands}.

Once deployed, there is no way to update a smart contract except for deleting it entirely and re-deploying a new one. Even a smart contract's author cannot modify the corresponding source code or fix bugs after that~\cite{antonopoulos2018mastering}. Due to such immutability, it is critical to ensure that a smart contract is secure before its deployment on a blockchain network, such as Ethereum. However, as usually happens to software, smart contracts suffer from vulnerabilities, which represent a risk as malicious attackers often exploit them. As an example, the DAO attack, in $2016$, resulted in a monetary loss of more than \$50 million dollars, which forced Ethereum to be hard forked and then rolled back to a previous state~\cite{mehar2019understanding}.

\begin{figure*}[t]
  \centering
  \includegraphics[width=1\textwidth]{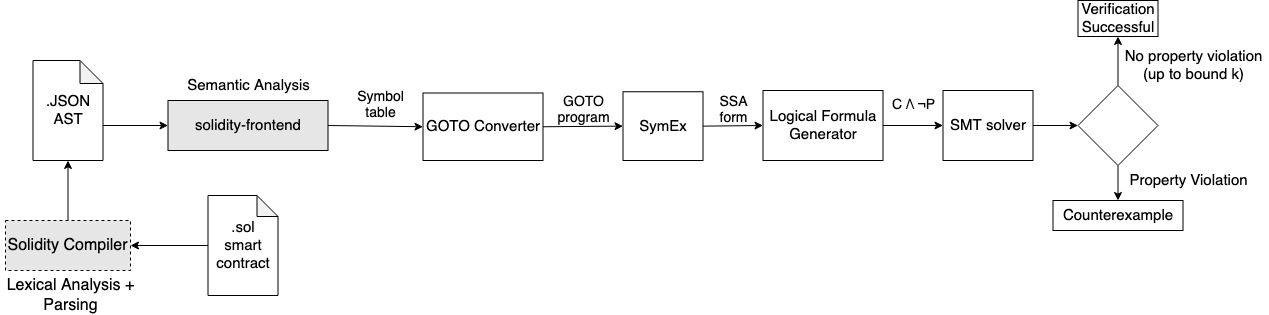}
\caption{Architectural overview of ESBMC with its extension for verifying Solidity smart contracts. 
}
\label{fig:new_verification_pipeline}
\end{figure*}


If there can be vulnerabilities, software testing becomes paramount, and the community has already begun to tackle the related problems~\cite{hajdu2019solc,garfatta2021survey}. However, most approaches target only a limited number of errors, which worsens as new applications appear and the need for specific aspects arises. Consequently, it is essential to employ mature and flexible verifiers, e.g., based on model checking and satisfiability modulo theories (SMT), to check smart contracts. This way, a myriad of problems is already handled, and consequent methodology advancements can then be devised faster, including behavior models and specific properties~\cite{cordeiro2020survey}.

In that sense, the efficient SMT-based context-bounded model checker (ESBMC) is a good candidate. It is a state-of-the-art (SOTA) checker, which can be extended to support different programming languages and target systems, such as digital filters and controllers, even incorporating behavior models and companion tools \cite{cordeiro2020survey,alhawi2021verification,chaves2019verifying}. In addition, it was initially devised as a C-language model checker and has been evaluated using standard benchmarks and embedded applications in the telecommunication industry~\cite{gadelha2018esbmc, cordeiro2011smt}. 

This paper tackles the problem raised here, i.e., smart-contract evaluation, and proposes a frontend for ESBMC based on the newly developed Grammar-based Hybrid Conversion methodology. 
It enables ESBMC to verify Solidity contracts 
via two steps. First, we convert Solidity JavaScript object notation (JSON) abstract syntax trees (AST) into the ESBMC's intermediate representation (IR). Second, we integrate with ESBMC's infrastructure 
to reuse its existing SMT-based verification strategies (incremental and \textit{k}-induction).

In order to evaluate the proposed framework, named ESBMC-Solidity, a benchmark suite with vulnerable smart contracts was created and used as input for it and other SOTA Solidity verification tools: Smartcheck~\cite{tikhomirov2018smartcheck}, Slither~\cite{feist2019slither}, Oyente~\cite{luu2016making}, and Mythril~\cite{consensysMythril}. As a result, ESBMC-Solidity outperformed the mentioned verification tools in soundness and performance. Besides, it identified all vulnerabilities, with a counterexample for each, and proved the fastest approach.

\section{Tool description}
\label{sec:tool}

\subsection{Tool Overview}
\label{section:tool-overview}

Fig.~\ref{fig:new_verification_pipeline} illustrates the architecture of ESBMC-Solidity, where the gray box with solid border represents the new frontend, and the white ones constitute the existing ESBMC's components. The gray box with a dashed border indicates an external element for preprocessing smart contracts: the Solidity compiler. It is used for lexical analysis and parsing, taking a smart contract as input and then transforming it into JSON AST, which is done with the argument \textit{-{}-ast-compact-json}.

The proposed approach takes JSON AST and converts each 
of its nodes into an equivalent IR one, using the ESBMC's \textit{irept}, a tree-structured IR that preserves a program's semantics.
Next, each \textit{irept} node is converted into the corresponding symbol and then added to a table, which is translated into a 
GOTO program. Then, the latter is processed by the symbolic execution engine (SymEx) to generate its static single assignment (SSA) form, which is used to generate verification condition (VCs) $C \land\neg P$, where $C$ represents constraints and $P$ denotes a safety property. Lastly, ESBMC uses off-the-shelf SMT solvers for verifying the satisfiability of those VCs. 

If a property is satisfiable, an execution path leads to a bug in an original Solidity smart contract. Then, when ESBMC detects it, a counterexample is provided, in the form of state traces, to allow its reproduction. It is worth noticing that ESBMC supports several SMT
solvers, including Z3, Bitwuzla, Boolector, MathSAT, CVC4, and Yices ~\cite{ESBMC_Github}. 

\subsection{The Grammar-Based Hybrid Conversion Methodology}
\label{section:grammar-based-hybrid-conversion}

Given an input smart contract, the goal of the proposed frontend is to populate the resulting symbol table, where each symbol is represented by the ESBMC's \textit{symbolt} data structure~\cite{ESBMC_Github}. Furthermore, it shall complete the type-checking procedure of Solidity AST nodes and transform each JSON AST node into its equivalent ESBMC's \textit{irept} one while preserving the associated semantic information. To achieve this goal, we developed this frontend based on the {Grammar-Based Hybrid Conversion} methodology, as an approach specifically devised for that during the development of this work.

\textbf{Grammar-Based Conversion}. The proposed frontend uses the library \textit{nlohmann/json}\footnote{JSON for Modern C++ - https://github.com/nlohmann/json} to process Solidity ASTs in JSON format. When traversing Solidity AST nodes, it uses different functions to transform them into equivalent $irep\_t$ ones, e.g., \textit{get\_var\_decl\_stmt}, \textit{get\_expr}, and \textit{get\_statement}, for \textit{variable-declaration-statement}, \textit{expression}, and \textit{statement} nodes, respectively. Besides, each AST node may contain multiple child ones, e.g., the AST node of a \textit{for} loop contains four child nodes: \textit{initialisation}, \textit{condition}, \textit{increment}, and \textit{loop body}. So, during their conversion, the production rules in Solidity grammar documentation\footnote{Solidity Grammar - https://docs.soliditylang.org/en/v0.8.6/grammar.html} are followed, so that they are visited in correct order. For instance, the variable-initialisation node of a \textit{for} loop must be visited before the \textit{body} one, as it may be referenced by the latter. If the node for a \textit{body} loop is converted before its variable initialisation, the type checker will fail to handle any reference to it. 

\textbf{Hybrid conversion}. Three functions must be supported: (1) $assert()$ for defining safety properties; (2) $assume()$ for defining constraints; and (3) $nondet()$ for assigning non-deterministic values to variables. Consequently, they are implemented by ESBMC as C-style declarations. However, the new frontend works with JSON AST nodes. Besides, since there are more than $70$ intrinsic declarations, e.g., forward declarations for nondeterministic types, we instantiate the existing ESBMC's clang frontend to convert those into \textit{irept} nodes to avoid replication, hence generating the symbol table mentioned before. Finally, the latter is further merged with the symbol table generated from the original Solidity AST. 

\subsection{Illustrative Example}
\label{section:illustrative-example}

\begin{figure} [!ht]
\centering
  \begin{subfigure}{.45\textwidth}
\begin{lstlisting}[escapechar=|]
pragma solidity >=0.4.26;
contract MyContract {
  uint8 x; |\label{contract:state_x}|
  uint8 sum; |\label{contract:state_sum}|
  function nondet() public pure
    returns(uint8) {
    uint8 i;
    return i;
  }
  function __ESBMC_assume(bool)
    internal pure { }
  function func_sat() external {
    x = 0;
    uint8 y = nondet();
    sum = x + y;
    __ESBMC_assume(y < 255); |\label{contract:start_of_constraint}|
    __ESBMC_assume(y > 220); |\label{contract:middle_of_constraint}|
    __ESBMC_assume(y != 224); |\label{contract:end_of_constraint}|
    assert(sum % 16 != 0); |\label{contract:property}|
  }
}\end{lstlisting}
\end{subfigure}
\caption{An example smart contract written in Solidity.}
\label{fig:illustrative_example_solidity}
\end{figure}

Fig.~\ref{fig:illustrative_example_solidity} shows an example of smart-contract verification with ESBMC-Solidity. 
Indeed, developers can 
instrument code, e.g., \textit{nondet} for nondeterministic integers between $0$ and $255$ and \textit{\_\_ESBMC\_assume} for additional constraints. Those help developers narrow down the scope for triggering a bug, hence identifying a set of breaking inputs. Function \textit{func\_sat} is the one we need to verify, where two state variables \textit{x} and \textit{sum} are defined in lines~\ref{contract:state_x} and~\ref{contract:state_sum}, respectively, while a safety property indicates that {\it x + y} should not be a multiple of $16$. In addition, constraints are added using \textit{\_\_ESBMC\_assume}, in lines ~\ref{contract:start_of_constraint}, \ref{contract:middle_of_constraint}, and~\ref{contract:end_of_constraint}, which restrict $y$ as any integer between $220$ and $255$, but $224$. Then, ESBMC will check whether there exists an execution path that satisfies its negation. This way, the verification of \textit{func\_sat} becomes a satisfiability problem: given the binary operation expression \textit{``sum = x + y''}, where \textit{x} is \textit{0} and \textit{y} is a constrained nondeterministic value, find an execution state where the negation of ``$sum \% 16 != 0$'' is satisfied. ESBMC is then invoked with 

\begin{verbatim}
esbmc <JSON AST> --function func_sat --z3.
\end{verbatim}

For the smart contract in Fig.~\ref{fig:illustrative_example_solidity}, ESBMC generates the $C$ and $P$ equations as described in~\eqref{C_eq:1} for constraints and~\eqref{C_eq:2} for property. Eq.~\eqref{C_eq:1} shows a conjunction of the constraints and assignments. When generating its SSA form, ESBMC uses the temporary variable {\it temp} to represent the left-hand-side of the safety property specified in line~\ref{contract:property}, which corresponds to the assignment $temp=sum\%16$ in Eq.~\ref{C_eq:1}. The resulting VC for satisfiability verification, via SMT solver, is then formed by $C\land\neg P$. Consequently, ESBMC reports a property violation and provides a counterexample that contains a trace of states showing the set of assignments and the breaking values that trigger such violation, where $y$ is set with a value $240$.

\noindent\begin{minipage}{.59\linewidth}
\begin{equation}
  C=\left[ \begin{array}{lr} y=nondet()\\ 
        \land\:sum=y\\
        \land\:y\,!=224\\
        \land\:temp=sum\%16 \end{array}\right] \label{C_eq:1}
\end{equation}
\end{minipage}%
\begin{minipage}{.45\linewidth}
\begin{equation}
 P=\left[ \begin{array}{lr} temp\,!=0 \end{array}\right] \label{C_eq:2}
\end{equation}
\end{minipage}

\section{Evaluation and Benchmarks}
\label{sec:exp}

\begin{table*}[t]
\begin{center} 
\begin{tabular}{|c|c|c|c|c|c|c|c|c|c|c|}
\hline 
\multirow{2}{*}{TC} &\multicolumn{2}{c|}{SmartCheck} &\multicolumn{2}{c|}{Slither} &\multicolumn{2}{c|}{Oyente} &\multicolumn{2}{c|}{Mythril} &\multicolumn{2}{c|}{ESBMC-Solidity}\\\cline{2-11} 
                    &Found  &CE   &Found  &CE  &Found  &CE                             &Found  &CE  &Found  &CE\\\hline
        TC1         &No     &-    &No     &-   &No     &-                              &Yes    &No  &Yes    &Yes\\\hline
        TC2         &No     &-    &No     &-   &No     &-                              &Yes    &No  &Yes    &Yes\\\hline
        TC3         &No     &-    &No     &-   &No     &-                              &Yes    &No  &Yes    &Yes\\\hline
        TC4         &No     &-    &No     &-   &No     &-                              &Yes    &No  &Yes    &Yes\\\hline
        TC5         &Yes    &N/A  &Yes    &N/A &\shortstack{Failed to \\ compile} &-   &Yes    &N/A &Yes    &N/A \\\hline
        TC6         &No     &-    &No     &-   &No     &-                              &Yes    &No  &Yes    &Yes\\\hline
        TC7         &No     &-    &No     &-   &No     &-                              &Yes    &No  &Yes    &Yes\\\hline
        TC8         &No     &-    &No     &-   &No     &-                              &Yes    &No  &Yes    &Yes\\\hline
        Total Time  &\multicolumn{2}{c|}{1.160s} &\multicolumn{2}{c|}{0.519s} &\multicolumn{2}{c|}{1.116s} &\multicolumn{2}{c|}{3.106s} &\multicolumn{2}{c|}{0.183s}\\\hline
\end{tabular}
\caption{Experimental results, where column ``Found'' indicates whether a bug was detected, followed by column ``CE'' showing whether a counterexample was provided. The line ``Total Time'' represents the CPU time~\cite{time_Linux_manual} used for verification.}
\label{table:Result}
\end{center}
\end{table*}

ESBMC-Solidity, though an early prototype, 
can detect vulnerabilities listed in the smart-contract weakness classification (SWC) registry~\cite{SWC_Registry}. So, our evaluation aims to answer three questions. \textbf{EQ1}: is our approach able to report a confirmed bug? 
(\textbf{soundness}). \textbf{EQ2}: does our approach find a bug in a reasonable amount of time? \textbf{(performance)}. \textbf{EQ3}: is our approach able to provide a counterexample to help reproduce a specific bug? \textbf{(bug reproduction)}.

\subsection{Benchmark Suite Design}
\label{sec:benchmark}

A benchmark suite that contains bugs in smart contracts was developed to evaluate ESBMC-Solidity and compare it with other SOTA verification tools. The design of each test case (TC) was guided by the SWC registry~\cite{SWC_Registry}, as shown in Table~\ref{table:TC_design}, while Table~\ref{table:Result} shows that all bugs in TCs were detected and confirmed by at least one of the non-ESBMC tools. The test suite and logs are publicly available in Zenodo\footnote{https://doi.org/10.5281/zenodo.5721726}.  
\begin{table}[!ht]
\begin{center}
\begin{tabular}{|c|c|c|}
\hline
SWC Bug ID                & Vulnerability                     & TC     \\ \hline
\multirow{2}{*}{SWC-101}  & Integer Overflow                  & TC1,2  \\ \cline{2-3}
                          & Integer Underflow                 & TC3,4  \\ \hline
SWC-115                   & Authorization through tx.origin   & TC5    \\ \hline
\multirow{2}{*}{SWC-110}  & Static array out-of-bounds        & TC6    \\ \cline{2-3}
                          & Dynamic array out-of-bounds       & TC7,8  \\ \hline
\end{tabular}
\end{center}
\caption{Test case design based on SWC registry~\cite{SWC_Registry}.}
\label{table:TC_design}
\end{table}

\subsection{Results}
\label{sec:results}

Table~\ref{table:Result} shows results for ESBMC-Solidity and other tools. The former 
found bugs in all TCs confirmed by Mythril, which affirms EQ1. SmartCheck and Slither were able to confirm TC5, which contains a vulnerability reported in the security considerations\footnote{https://docs.soliditylang.org/en/v0.8.6/security-considerations.html}, while detected none for the other TCs. Oyente did not find any bug. Mythril, a tool used in the service MythX\textsuperscript{TM}~\cite{consensysMythril}, also reported bugs in all TCs. 

Apart from TC5, Mythril and the other non-ESBMC tools failed to provide a counterexample for each TC. However, ESBMC-Solidity did, which affirms EQ3. For TC5, a counterexample is not needed, as a tool should only inform that authorization via $tx.origin$ must be avoided. One possible reason for the missing counterexamples could be loss of the original Solidity syntax, as tools either use EVM bytecode, e.g., Oyente and Mythril or rely on various forms of IR that do not preserve the original Solidity declaration references needed for state tracing, e.g., SmartCheck and Slither.

ESBMC-Solidity is the fastest tool, as can be seen in the last line of Table~\ref{table:Result}, which thus affirms EQ2. Oyente and Mythril work on EVM bytecode and employ simulation for execution path-exploration~\cite{alt2018smt}, which might be the reason why they are slower than ESBMC-Solidity. Apart from bug detection, Slither also provides code optimization~\cite{feist2019slither}, which might add to the total verification time. To the best of our knowledge, there is no option in Slither to disable the optimization. SmartCheck is implemented in Java, converts Solidity code into an XML-based IR, and uses XPath to query it, while ESBMC-Solidity is implemented purely in C++. 


\section{Related Work}
\label{sec:related_work}

Among the tools we evaluated, Mythril and Oyente use SMT-based symbolic execution to check EVM bytecode and also simulate a virtual machine for execution-path exploration, which might lead to performance degradation~\cite{alt2018smt}. ESBMC-Solidity also uses SMT solvers as backends, but it processes ASTs, so there is no need for environment simulation. 

A 
tool that also adopts SMT encoding and solvers to find satisfiability for a property violation is discussed by Alt and Reitwiessner~\cite{alt2018smt}. They developed a component to translate programs into smtlib2 formulae to interface with SMT solvers~\cite{alt2018smt}. 
The main difference between it and ESBMC-Solidity is that the latter supports code instrumentation, 
which narrows down the scope of inputs that trigger violations. In addition, ESBMC-Solidity can also be extended on top of various existing verification strategies and reasoning techniques provided by ESBMC, such as \textit{k}-induction. 

\section{Conclusions}
\label{sec:conc}

We presented ESBMC-Solidity that checks memory safety and user-defined properties in smart contracts written in the Solidity programming language. We evaluated ESBMC-Solidity against other SOTA verification tools and overcame them, confirming all presented bugs and providing the associated counterexamples. Although ESBMC-Solidity is an early prototype, it shows promising results. Our current focus is on providing $100$\% coverage for the language Solidity, including polymorphism, inheritance, special crypto functions, such as Keccak256 and sha256, and multiple returns. 


\bibliographystyle{ACM-Reference-Format}
\bibliography{toolpaper}

\end{document}